# Dynamic Viscosity of Methane Hydrate Systems from Non-Einsteinian, Plasma-Functionalized Carbon Nanotube Nanofluids

*Adam McElligott, André Guerra, Chong Yang Du, Alejandro D. Rey, Jean-Luc Meunier, Phillip Servio\**

Department of Chemical Engineering, McGill University, Montreal, Quebec H3A 0C5, Canada

*phillip.servio@mcgill.ca

ABSTRACT

The viscosity of oxygen-functionalized multi-walled carbon nanotube (O-MWCNT) nanofluids was measured for concentrations from 0.1 to 10 ppm under conditions of 0 to 30 MPag pressures and 0 to 10 °C temperatures. The presence of O-MWCNTs did not affect the temperature dependence of viscosity but did reduce the effective viscosity of solution due to cumulative hydrogen bond-disrupting surface effects, which overcame internal drag forces. O-MWCNTs added a weak pressure dependence to the viscosity of solution because of their ability to align more with the flow direction as pressure increased. In the liquid to hydrate phase transition, the times to reach the maximum viscosity were faster in O-MWCNT systems compared to the pure water baseline. However, the presence of O-MWCNTs limited the conditions at which hydrates formed as increased nanoparticle collisions in those systems inhibited the formation of critical clusters of hydrate nuclei. The times to viscosity values most relevant to technological applications were minimally 28.02 % (200 mPa·s) and 21.08 % (500 mPa·s) slower than the baseline, both in the 1 ppm system, even though all systems were faster to the final viscosity. This was attributed to O-MWCNT entanglement, which resulted in a hydrate slurry occurring at lower viscosity values.

1. INTRODUCTION

Gas or clathrate hydrates are increasingly being examined for novel natural gas transport applications and carbon capture technologies such as carbon dioxide sequestration and separation processes, including flue gas treatment and fruit juice concentration.[1-6] Hydrates are a class of crystalline inclusion compounds that typically arise under high (megapascal range) pressures and moderate temperatures. They consist of (1) a gas or volatile liquid that becomes physically trapped in (2) a host lattice of water molecule hydrogen bonds. During the phase transition, no reaction or chemical bonding occurs. Instead, weak van der Waals forces between the guest molecule and the crystal lattice confer stability over the entire crystal structure. The size of the enclosed molecule determines which type of hydrate structure forms. These are structure I, for smaller gasses like methane or carbon dioxide, structure II, for larger molecules like propane or tetrahydrofuran, and structure H, which enclathrate larger guests (up to 0.9 nm in diameter).[1] Additionally, not every hydrate cage must be occupied for the structure to remain stable, so hydrates also have non-stoichiometric properties.[7]

There are generally three distinct stages to hydrate formation in a well-mixed system: dissolution, induction, and growth. These are outlined in **Figure 1** in the context of a gas consumption curve. In the dissolution stage, the hydrate-forming inclusion molecule dissolves into the liquid phase until it reaches a saturation point ($n_{sat}$, its equilibrium solubility) at $t_{sat}$. As dissolution continues beyond this time, the solution becomes supersaturated, which is necessary for hydrate formation to occur. The difference between $n_{sat}$ and the current liquid saturation, $n_{liq}$, is called the degree of supersaturation and is a crucial indicator of the magnitude of the driving force for hydrate formation. The supersaturated period is called the induction stage, where the molecules in the system form microscopic clusters of hydrate nuclei to reduce the system's free energy. However, they dissociate continually until a critical cluster radius is attained and the nucleus reaches energetic stability.[1] The critical nucleus size can range from 30 to 170 Å for methane hydrates.[8] The formation of this thermodynamically stable nucleus occurs stochastically at the nucleation point, $t_{nuc}$, marking the spontaneous onset of autocatalytic, exothermic hydrate growth. This is the growth stage, where the consumption of inclusion bodies initially increases linearly with time though eventually slows either due to insufficient water to form more cages or low supply of the guest molecule.[1] In fact, growth can be halted, and a closed system can return to equilibrium, if the supersaturated regime (the driving force for formation) is not maintained.[8] All of these stages are examined in this study, though the driving forces for formation were often sufficiently great that the dissolution and induction stages were extremely quick; they sometimes occurred in just a fraction of a second. The primary focus of this study will thus be hydrate formation from the start of growth onward.

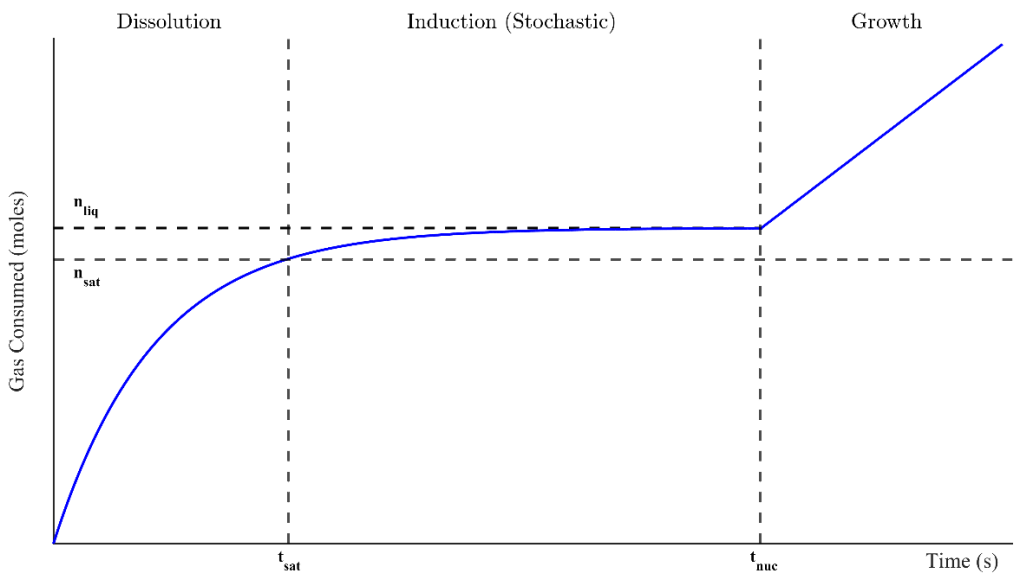

**Figure 1.** Schematic of a typical hydrate formation profile through guest gas consumption. Here, $n_{sat}$ is the molar saturation value that occurs at $t_{sat}$ and $n_{liq}$ is the final number of moles in the supersaturated solution at nucleation, which occurs at $t_{nuc}$.

Hydrates have high gas storage potentials and form under less severe conditions to form transportable compounds. For instance, methane hydrates form at much higher temperatures than liquified and lower pressures than compressed natural gas.[9] Additionally, they have highly selective formation properties and efficient structural packing. These make them ideal for novel industrial applications.[2, 4-6] In many of these novel processes, there are often heat and mass transfer limitations that can be overcome through the addition of optimizing additives. The promotional effects of these additives, usually surfactants or nanoparticles, can be kinetic or thermodynamic in nature. The former is when the additive induces the nucleation of stable hydrate clusters while the latter shifts the three-phase equilibrium curve towards increasingly milder formation conditions.[10] Nanoparticle additives often include metal oxides or carbon-based nanoparticles like graphene nanoflakes or carbon nanotubes.[11-13] Notably, hydrate yields in aqueous multi-walled carbon nanotube (MWCNT) systems have been measured to be 4.5 times higher than in pure water.[14, 15] However, these nanoparticles are made of carbon and naturally hydrophobic, so they agglomerate in and settle out of aqueous solutions.[14] Therefore, they are often mixed with surfactants or chemically treated to maintain their stability.[16] Current advances allow for the addition of oxygen or amine functional groups to the MWCNT surface via plasma treatment processes. Adding covalently bonded oxygenated functionalities like hydroxyl or carboxyl groups creates hydrophilic O-MWCNTs (oxygen-functionalized MWCNTs) that remain dispersed in aqueous solution for at least two years.[16] O-MWCNTs have been found to increase the dissolution rates of methane in water as well as the growth rates of methane hydrates, both maximally by approximatley 16% compared to pure water.[17, 18] However, the viability of hydrate technologies does not solely depend on promotion as many of these technologies propose to use semi-batch or continuous processes for operation: a flow state will be maintained while hydrates are forming. Therefore, significant control of the system viscosity will be necessary to avoid complete solidification and reduce pumping requirements. A rheological characterization of O-MWCNT nanofluids under different thermodynamic conditions during the liquid-to-hydrate phase transition is critical to guiding the design of these optimized processes.

This study will characterize the shear rheology of methane hydrate formation in O-MWCNT nanofluids systems with temperatures from 0 to 10 °C and pressures from 0 to 30 MPag. Previous studies have examined hydrate systems from pure water and plasma-functionalized graphene nanoflake nanofluids under the same thermodynamic conditions, and there have been many recent studies on the viscosity of nanotube nanofluids, primarily in ethylene glycol or water.[10, 19-29] To the best of our knowledge, this is the first time the viscosity of plasma-functionalized carbon nanotubes has been measured in any system, including liquid and hydrate systems. There have been few high-pressure nanotube-based nanofluid viscosity studies, mainly examining their rheological behaviour in drilling muds.[30-32] However, these studies do not examine the range of temperatures, pressures, or concentrations useful for applications based on gas hydrates. Moreover, though other nanofluids have been analyzed at similar pressures, evaluating the viscosity of nanotube nanofluids at high pressure in pure water is entirely novel.

## 2. MATERIALS AND METHODS

### 2.1 Experimental Setup

Details of the experimental setup can be found in previous studies.[29] Briefly, the measurement device was an Anton Paar MCR 302 rheometer equipped with a high-pressure cell (maximum 40 MPag) into which the double-gap measurement geometry was inserted. The sample could be loaded into the double annulus space, and rotational motion was induced in the geometry via a magnetic measurement head. The sample temperature was maintained by a Julabo F-32 chiller using a 50/50 mixture of ethylene glycol and water. The methane used in the experiments was purchased from MEGS and was of ultra-high (99.99%) purity, while the reverse osmosis (RO) water had a maximum organic content of 10 ppb. Pressures up to 10 MPag were achieved solely using a gas cylinder, though higher pressures required a Schlumberger DBR high-pressure positive displacement pump to compress methane gas samples (in its piston chamber) and reach the required pressures for this study. The O-MWCNTs used in this study were produced in McGill University's Plasma Processing Laboratory and are described in the following section.

2.2 Characteristics of Oxygen-Functionalized MWCNTs

The O-MWCNTs described herein were produced and characterized in McGill University's Plasma Processing Laboratory prior to this study. The fabrication process will be provided briefly, focussing on those aspects relevant to the final characteristics of the nanoparticles, and having the greatest influence on the effects the O-MWCNTs engender in the system. O-MWCNTs are produced in two stages. The first uses a chemical vapour deposition process wherein the growth of as-produced MWCNTs, with acetylene as a carbon source, occurs on stainless steel meshes. In the second stage, the MWCNTs are exposed to a capacitively-coupled radio frequency glow discharge in a mixture of argon, ethane, and oxygen. This adds covalently bonded, hydrophilic oxygenated functional groups to the surface, such as hydroxyl, carboxyl, and carbonyl groups. The diameter of the functionalized O-MWCNTs is approximately 30 nm on average, and their lengths are approximately 10 μm. The atomic composition of the surface is mostly carbon, though about 21 % of the surface is oxygen. Through ultrasonication, the O-MWCNTs are harvested from the growth surface and dispersed in RO water to produce a homogeneous nanofluid. Oxygen functionalization above 14% showed full nanofluid stability over several years without any required surfactant.[16] Further information regarding the production, functionalization, characterization, and imaging of the O-MWCNTs is found in Hordy et al. (2013).

2.3 Experimental Procedure

Initially, 7.5 mL of the O-MWCNT solution was loaded into the high-pressure cell. Then, the measurement geometry was inserted and used to close the cell. Using methane at a pressure of 1 MPag, the sample's headspace was purged five times to eliminate any air. Once the sample temperature was stable (within 0.1 °C of the setpoint), the measurement system was activated. The cell was then instantly charged with methane gas which came directly from a gas cylinder or the piston chamber of the positive displacement pump. The rheometer ran at a constant 400 $s^{-1}$ shear rate, the recommended shear rate for double-gap measurement geometries if the samples are low viscosity liquids. The shear rate would remain constant throughout a run so that consistent and comparable temporal viscosity measurements could be obtained during the different stages of hydrate formation. Therefore, the Newtonian nature of O-MWCNT nanofluids is out of the scope of this study as shear rate changes would be required to determine if such effects were present.

The pressures examined in this study ranged from 0 to 5 MPag (going up by 1 MPa) and 10 to 30 MPag (going up by 5 MPa). The temperatures examined ranged from 0 to 10 °C (going up by 2 °C). Together, these account for 66 different pressure/temperature combinations and are the same conditions as Guerra et al. (2022), who examined methane hydrate formation in pure water systems, which will be used as a baseline for comparison.[29] The majority of these conditions are found above the three-phase equilibrium line and so are classified as hydrate-forming: there is a positive driving force for hydrate formation.[33] All these runs were given a 90-minute period to begin formation, which was always detected by a rapid increase in viscosity. Viscosity measurements would continue until the rheometer's set maximum torque limit was reached. This would occur a few seconds before complete solidification of the sample, as solidifying with every run could significantly damage the ball bearings in the magnetic measurement head. At this time, the rheometer would stop collecting data automatically, and the run would end. The Anton Paar software RheoCompass v.1.25 was used for data collection, and data analysis was performed in MATLAB®. For those conditions which were not expected to form hydrates, the measurement progressed until a continuous ten-minute period of stable viscosity, usually within ±0.005 mPa·s. This was to ensure that effects from gas dissolution or temperature changes at the start of the run were eliminated from the final value. The concentrations of the O-MWCNT solutions were 0.1, 1, and 10 ppm by mass, and all 66 conditions were tested for each concentration. Additionally, the viscosity of plasma-functionalized nanotube nanofluids has never been measured, so unpressurized runs of O-MWCNT nanofluids with concentrations of 0.1, 0.5, 1, 5, and 10 ppm by mass were also examined with no methane present (i.e., in an air atmosphere). This concentration range is rarely investigated in nanofluid viscosity studies and has never been investigated for nanotube nanofluids, regardless of functionalization.

## 3. RESULTS AND DISCUSSION

### 3.1 Viscosity of O-MWCNT Nanofluids

Before examining the solution viscosity during the formation of hydrates in the presence of O-MWCNT nanofluids, it was necessary to measure the unpressurized viscosity of the nanofluid without methane present. This was to provide a reference to which the methane-pressurized results could be compared, as the viscosity of plasma-functionalized carbon nanofluids and the effects present in these systems have never been measured. Therefore, viscosity was measured across the concentration range (0.1, 0.5, 1, 5, and 10 ppm) for each temperature condition (0, 2, 4, 6, 8, and 10 °C), and the effects of concentration and temperature on viscosity, as well as relative viscosity compared to pure water, were investigated.

The effects of temperature on O-MWCNT nanofluid viscosity are presented in **Figure 2**. The absolute viscosity of the solution decreases linearly with increasing temperature for all concentrations. This behaviour is commonly observed in nanofluids over even larger temperature and concentration ranges than those present in this study and has been observed in several recent nanotube nanofluid studies as well.[19-21, 23-27] Furthermore, all linear decreases appear to be similar to that of the baseline, which may be because nanofluids are often measured not to have a significant effect on the temperature dependence of the base fluid viscosity.[34-37] Instead, more energy is supplied to the fluid molecules as temperature increases, reducing intermolecular adhesion forces and thus the fluid's resistance to shear and viscosity.[34, 38, 39] Note that certain nanoparticles can exhibit agglomeration as temperature increases. However, it is uncommon with functionalized nanoparticles like O-MWCNTs and was not measured here.[16] Therefore, agglomeration is not expected to have a role in the effective viscosity of the system. To confirm that O-MWCNTs do not affect the temperature-viscosity relationship, the viscosity of the nanofluid divided by that of the base fluid under the same conditions (the relative viscosity) can be observed in **Figure 2**b. As the temperature increased, no significant change in relative viscosity was observed. It may appear that the relative viscosity decreases with temperature at the lower concentrations, but it is not believed that these changes are significant, and most of them are within the error of the measurement device. McElligott et al. (submitted) measured the viscosity of plasma-functionalized graphene under the same conditions and concentrations and determined no significant effect.[40] Therefore, as the system temperature increases, a solution with carbon nanoparticles can be treated as some dispersion in a progressively lower-density liquid.

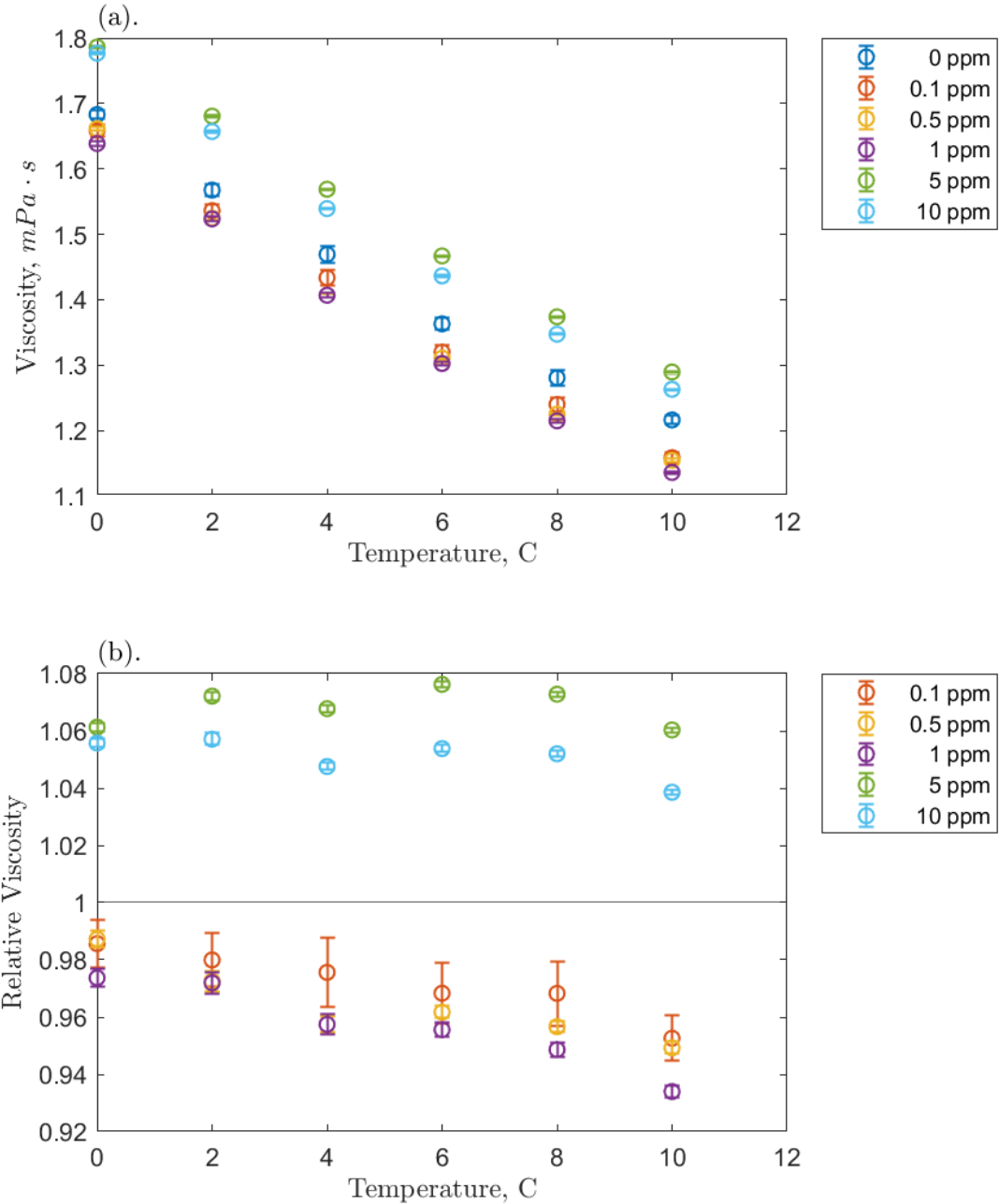

**Figure 2.** Temperature effect on the measured viscosity of the O-MWCNT-water system studied. This includes (a) the absolute viscosity and (b) the relative viscosity compared to a pure water baseline. Bars represent the 95% confidence intervals on the data, which are very small.

**Figure 2**b demonstrates a significant effect of O-MWCNTs in the solution: the relative viscosity at lower concentrations (from 0.1 to 1 ppm) has a value of less than one. Contrary to nearly all nanofluid viscosity studies, including nanotube nanofluids, this may indicate that the addition of O-MWCNTs could reduce the effective viscosity of the system.[10, 19-27, 34] This behaviour is termed non-Einsteinian (NE) viscosity as Einstein was first to develop equations showing that the addition of solid particles to a fluid system increases the internal friction coefficient (intra-solution drag) and thus the fluid effective viscosity.[41] Viscosity increases can occur not only through the hindrance of fluid particle motion but also from nanoparticle-liquid interactions, which may create interacting electrical double layers on the suspended particles and add new electroviscous forces to the solution.[25, 38] However, NE behaviour has been reported in glycol-based metal oxide nanofluids, where it was suggested that the nanoparticle disrupted the hydrogen bonds upon which glycol viscosity strongly depends.[42] This said, the viscosity of water also depends on hydrogen bond strength, though NE behaviour is very rarely reported in water-based nanofluids. Those studies that report this behaviour consistently involve low concentrations of dispersed nanotubes, like in the present study.[28, 30, 43] Specifically, the behaviour was attributed to the "lubricative effect"[28] of multi-walled carbon nanotubes (MWCNTs), though further elucidation is required.

**Figure 3** breaks down a hypothesis of NE viscosity in water, first proposed in detail by McElligott et al. (submitted), though it should be noted that the effects are on the molecular scale and thus were not tested as part of this study.[40] Computational modelling would be required to test this hypothesis and is therefore suggested as immediate future work. At ultra-low concentrations such as those in this study, surface effects from dispersed nanofluids (with high specific surface areas) may overcome internal friction effects, which would otherwise increase the effective viscosity of the solution. For instance, like in **Figure 3**a, a solvation layer that develops at the hydrophobic portion of the O-MWCNT surface (approximately 79% of the surface) could weaken the local hydrogen bonding structure.[35] Shelton (2011), for example, has shown that enhanced density fluctuations at nanoparticle surfaces induce increases in the local free volume.[44] Therefore, there may be weaker intermolecular attraction forces and more empty diffusion sites available for water molecules at the solid-liquid interface. The high specific surface area of O-MWCNTs, which are well-dispersed due to oxygen functional groups, may augment the free volume sufficiently to overcome drag forces. Upon examination of previous nanotube studies, it is notable to observe that NE behaviour also occurred only at low concentrations in all cases. In these studies, the MWCNTs were also functionalized, or a dispersant was used.[28, 30, 43] Therefore, the sum of local density reduction effects may be greater than the small amount of drag present at low concentrations as the surface area on which these effects occur remains high. Additionally, local stresses generated by shearing the fluid create large density fluctuations. These fluctuations are akin to acoustic waves and so they travel faster through solids than liquids. Therefore, the presence of solid particles may enhance wave propagation. Notably, amplified acoustic waves can improve momentum transport through particle-water collisions. As illustrated in **Figure 3**b, a wave in the liquid state with a certain velocity before contact with a nanoparticle may have a much greater velocity when the same wave returns to the liquid state. The flux in the shear direction would be more significant, there would be a smaller impedance to flow, and viscosity would effectively decrease.[44]

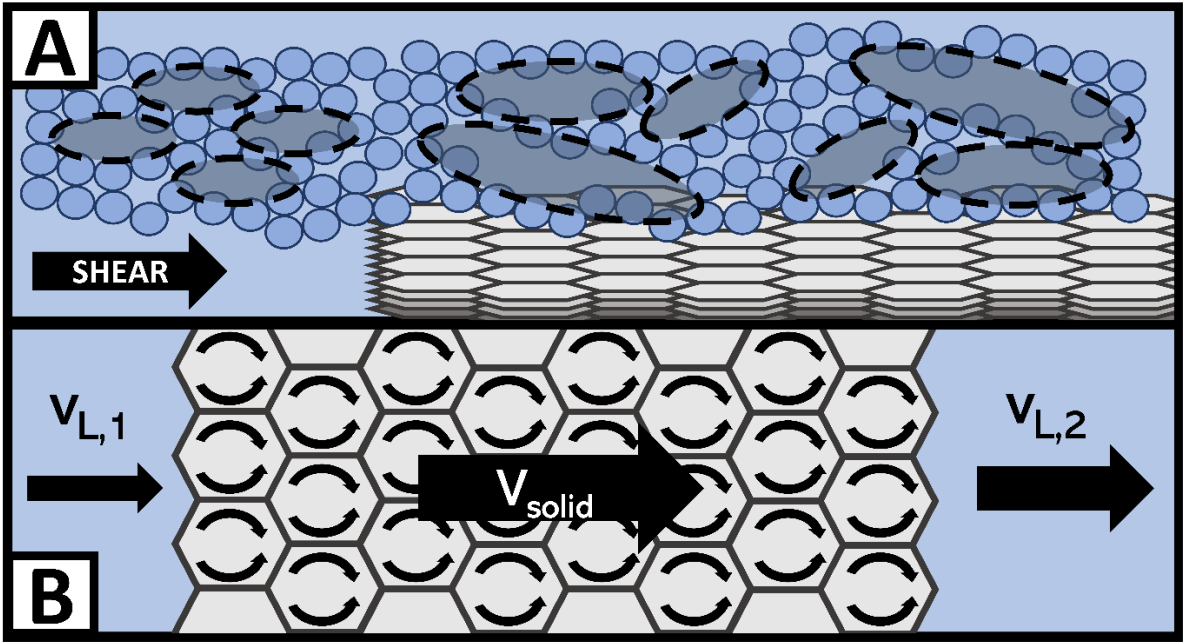

**Figure 3.** Examples of non-Einsteinian viscosity effects at the O-MWCNT surface include (A) increases in free volume, where the small blue circles are water molecules and the grey ovals are free volume cavities, and (B) density wave enhancement through the carbon structure with crossflow.

Returning to **Figure 2**b, the NE relative viscosity is consistent from 0.1 to 1 ppm with values between 0.94 and 1. However, at the higher concentrations (5 and 10 ppm), there is a transition to the Einsteinian viscosity regime, and the relative viscosity ranges from 1.04 to 1.08. The Krieger-Dougherty equation is often used to predict the relative viscosity using concentration as a parameter. The equation is as follows:

$$\eta_r = \frac{1}{\left(1 - \frac{\phi}{\varphi_m}\right)^{\mu \varphi_m}} \qquad \text{Equation 1}$$

Where $\eta_r$ is the relative viscosity, $\phi$ is the mass fraction in % assuming no agglomeration (an appropriate assumption for this system), $\varphi_m$ is the maximum particle packing fraction assumed to be 0.62,[45] and $\mu$ is the intrinsic viscosity which is approximately 2.5 for monodispersed solutions.[46] The minimum value this equation can output is 1.00 and, therefore, it cannot be used to evaluate the NE behaviour between 0.1 and 1 ppm. However, at 5 and 10 ppm (0.05 and 0.1 wt.%), it predicts relative viscosities of 1.14 and 1.31, respectively, compared to the average measured values of approximately 1.06 and 1.07. This overprediction suggests that NE effects could still be present to a significant degree at these loadings and that the Krieger-Dougherty equation may not accurately capture the relative viscosity behaviour of nanoparticles with high specific surface areas at low concentrations. Furthermore, McElligott et al. (submitted) found that for oxygen-functionalized graphene nanoflakes (O-GNFs) in the same temperature and concentration range, the relative viscosity remained non-Einsteinian (between 0.96 and 0.99) even at higher concentrations.[40] This is despite evidence that platelet-shaped nanoflake structures result in greater internal friction coefficients.[10] Both O-GNF and O-MCWNT particles are well-dispersed, but graphene often has more than double the specific surface area of carbon nanotubes.[47] If NE behaviour is predominantly the result of surface effects, then it follows that a particle with more available surface area for the same mass percent will show more NE effects. Drag caused by geometric effects is further overcome by a greater accumulation of surface viscosity reductions at higher concentrations.

### 3.2 Dynamic Viscosity of Methane Hydrate and O-MWCNT Systems

#### 3.2.1 Temperature and Pressure Effects

The effects of temperature and pressure were measured in the same temperature range for concentrations of 0.1, 1, and 10 ppm O-MWCNT. **Figure 4** shows the effects using both isobars and isotherms for the 1 ppm O-MWCNT system. The same figures for the 0.1 and 10 ppm systems are in the Appendix. Each test condition averages 110 points taken over 10 minutes at a constant shear rate. Hydrate formation occurred near-instantaneously in systems with pressures of 10 MPag and above, so it was not possible to measure constant liquid viscosity values at these pressures; dynamic viscosity profiles are given in section 3.2.2. Therefore, only the measurements from 0 to 5 MPag are presented in **Figure 4**. Note that this smaller pressure range should also include eight conditions under which hydrate formation is favourable. However, hydrates did not form under these conditions: this will be discussed later in section 3.2.2.

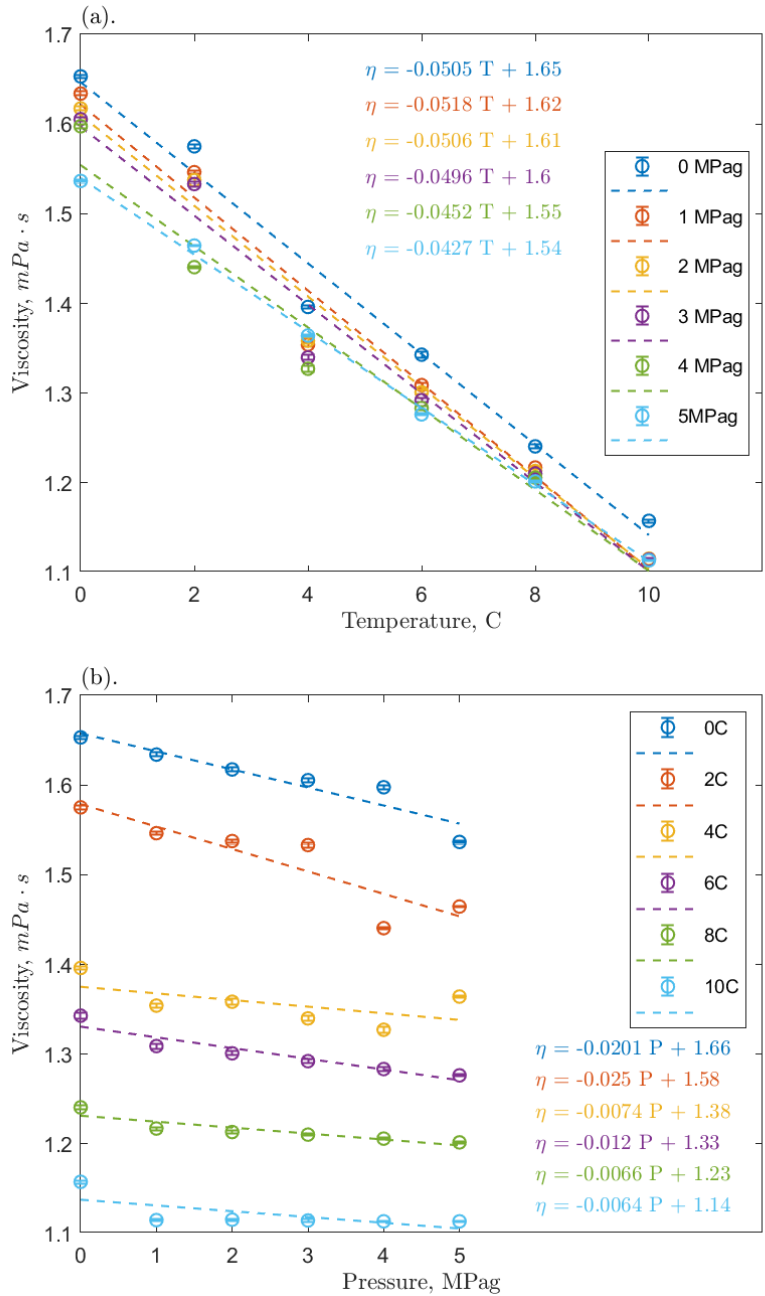

**Figure 4.** Effect on the measured viscosity of the 1 ppm O-MWCNT-methane-water system of (a) temperature and (b) pressure; error bars represent the 95% confidence intervals on the measured mean viscosity.

Linear regressions were implemented to analyze the effects of temperature and pressure across the range. **Figure 4**a shows that the nanofluid viscosity again decreased with increasing temperature. All slopes are reasonably similar across pressures, the average slopes were -0.0352, -0.0484, and -0.0342 mPa·s/°C for the 0.1, 1, and 10 ppm concentrations, respectively, compared to -0.0445 mPa·s/°C for pure water/methane.[29] These slopes demonstrate a clear negative trend that is relatively consistent regardless of the addition of nanoparticles and indicates that O-MWCNTs likely do not affect how the viscosity of water changes with temperature in this system. This conclusion agrees with other high-pressure nanofluid viscosity studies, including one using O-GNFs, which determined that viscosity changes with temperature depended only on the base fluid.[32, 48] Note that the average slopes at the 0.1 and 10 ppm loadings are approximately 30% smaller than that of the baseline, which could indicate that the viscosity is more stable at these concentrations (i.e., the temperature-dependence of viscosity is reduced). However, as the slopes (1) are relatively small, (2) are within 0.01 mPa·s/°C of the baseline, and (3) remain in the same order of magnitude, further investigation on a significantly greater range of temperatures would thus be required to form a strong conclusion. Additionally, the Vogel-Fulcher-Tammann-Hesse (VFT) or Arrhenius-type equations could be used to characterize the temperature dependence of nanofluid viscosity here. However, in smaller temperature ranges, such as the one present in this study, using these non-linear models could result in overfitting, and linear regressions are sufficient.

**Figure 4**b shows how nanofluid viscosity changed with pressure. Again, all slopes were reasonably similar across temperatures, and the average slopes were -0.01782, -0.01292, and -0.00685 mPa·s/MPag for the 0.1, 1, and 10 ppm concentrations, respectively, compared to 0.00125 mPa·s/MPag for pure water/methane. It could be said that these slopes are all reasonably close to zero and are consistent regardless of the addition of nanoparticles. Notably, however, the average slopes at 0.1 and 1 ppm are negative and an order of magnitude greater than that for the baseline,

indicating a weak pressure effect. Hydrophobic methane molecules may adsorb onto the hydrophobic portion of the O-MWCNT surface to reduce the excess Gibbs free energy of the system.[49] These molecules could then be shuttled into the liquid bulk, slightly increasing the presence of methane bubbles in solution, which would result in a minor viscosity reduction effect.[29] It is notable that in all cases, the average value of the slope is most strongly influenced by higher magnitude slopes at lower temperatures, where the solubility of methane is highest, even though it is still sparingly soluble. This could indicate that the presence of more methane strongly influences the pressure-dependent viscosity of solution. Previous studies have suggested that O-MWCNTs may not significantly change the amount of methane in aqueous systems.[18] However, NE behaviour depends on the accumulation of minor surface effects, so it is possible that even small amounts of methane could significantly impact the effective viscosity.

Furthermore, McElligott et al. (submitted) found a weak pressure dependence only at 10 ppm in O-GNF systems, though the specific surface area is more significant in that system and, therefore, there would be a greater presence of methane.[40] In other words, if the increased presence of methane was the only effect leading to a pressure-dependent viscosity, the O-GNF systems, like the O-MWCNT systems, should exhibit such a dependence at all concentrations. It may be that pressure increases affect the conformation of MWCNTs to reduce viscosity further. High aspect ratio nanostructures have been shown to align with the flow direction in sheared fluids.[34] O-MWCNTs are relatively long (about 10 μm), and the more their length aligns with the flow path, the less drag they impose on the system. Pressure increases may result in greater nanotube alignment (conformation to the flow direction), leading to additional decreases in viscosity, regardless of what gas is used to impose that pressure. Moreover, this could explain why the pressure effects are greater at lower temperatures: in higher density water, there is less mobility for the orientation of the MWCNTs, and they may be more likely to have higher degrees of alignment. These alignment effects would not occur in O-GNF systems as those nanoparticles are much more rigid (only 100 nm in length). Significant pressure effects on their conformation would thus not be expected, and any pressure-dependent viscosity effects rely on more methane in the system. Therefore, the geometry of the O-MWCNTs could allow for a weak pressure dependence of viscosity at all concentrations via pressure-dependent conformational changes, which are more substantial than surface-area dependent methane bubble additions.

The effects of temperature and pressure on the relative viscosity of the solution were also examined for each concentration and are presented in **Figure 5**. Note that because it was determined that there was only a weak pressure effect on viscosity, the values at each temperature/concentration combination are averaged over the 0 to 5 MPag pressure range. Additionally, the base fluid viscosity is not pure water, but water pressurized with methane in the same pressure range. This eliminates the influence of methane, which, as mentioned, lowers the viscosity of the solution.

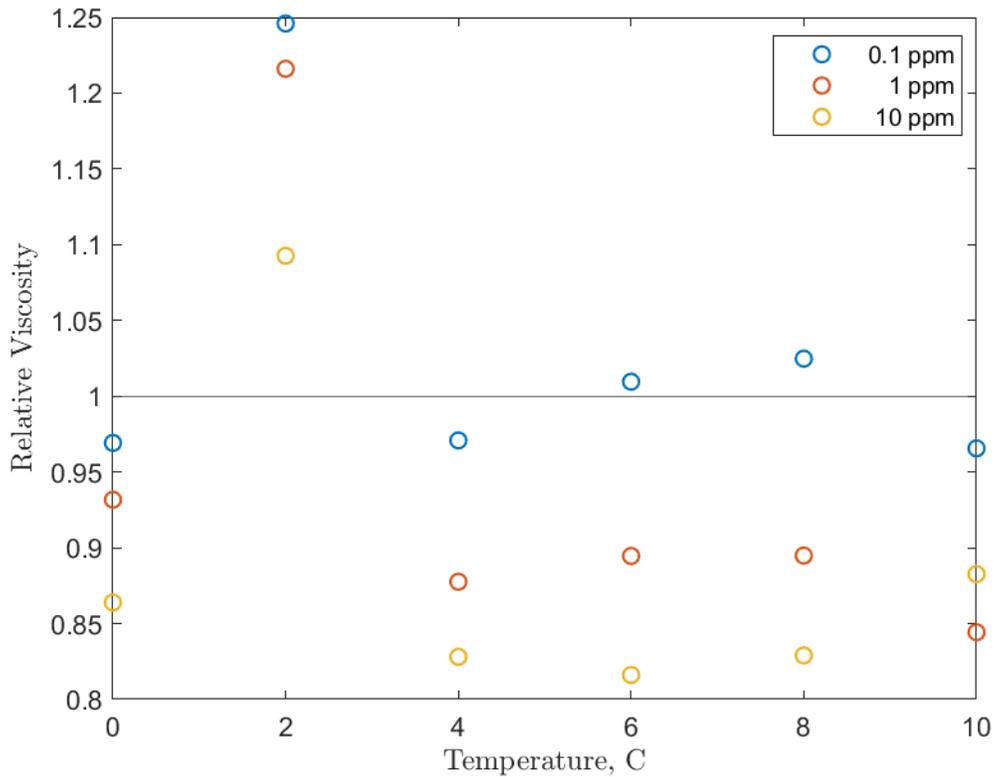

**Figure 5.** Temperature effect on the relative viscosity of the O-MWCNT-methane-water system studied with values averaged over the 0 to 5 MPag range.

Examining **Figure 5**, from 4 to 10 °C, the relative viscosities are nearly equal for each concentration and are largely non-Einsteinian. As concentration increases, the ranges of variation of the relative viscosity values with temperature decrease: 0.95 to 1.05 for 0.1 ppm, 0.85 to 0.95 for 1 ppm, and 0.8 to 0.9 for 10 ppm. These ranges are lower or within the relative viscosity ranges observed in unpressurized air tests (0.93 to 1). Notably, the trend is reversed from the unpressurized tests: the lowest relative viscosity is at 10 ppm (which is now also non-Einsteinian). As concentration rises, the available surface area for gas adsorption rises. Additionally, conformational alignment effects could be greater at higher loadings with more particles to align. Therefore, the results increase the likelihood that the nanoparticles are (1) bringing additional methane into the system and (2) aligning more with the flow direction under pressure, which would further lower the effective viscosity and may provide a small pressure-dependence to that viscosity. In short, increases in concentration would also increase internal friction, but the further strengthened viscosity-reducing effects related to pressure overcome these increases, and relative viscosity instead decreases with concentration. This is also a reverse of the trend found in O-GNF systems, where the relative viscosity increased with concentration. However, major conformational effects are not present in those systems, so there are generally fewer viscosity reduction effects to counteract increases in internal friction or shorter mean free paths.

There is also a significant rise in relative viscosity at 2 °C in **Figure 5**, well into the Einsteinian regime, which may be related to the density anomaly of liquid water. This rise occurred at the same temperature and was of the same magnitude (0.3) in the O-GNF systems, indicating that the rise is closely related to both the characteristics of water and the presence of the nanoparticles. It is expected that density increases as temperature decreases, but water has a density maximum at 4 °C. It has previously been suggested that this is a transition between hydrogen bond orderings where, below 4 °C, the bond ordering becomes more tetrahedral, so closer to the structure of ice, which is of lower density than liquid water.[50] However, water with ice-like structures has been measured to have a higher viscosity, so a relative viscosity increase would also be expected during this transition, though it occurs at 2 and not 4 °C. Decreases in hydrogen bond strength of less than 2% can shift the density anomaly towards colder temperatures, and, as O-MWCNTs may reduce the strength of the h-bond network, the increase in relative viscosity from the presence of ice-like structures occurs at 2 °C.[51] Furthermore, while more ice-like structures should increase the relative viscosity at 0 °C, there is instead a return to NE values similar to what they were from 4 to 10 °C. While more ice-like patches could be present at 0 °C, the shear rate may be large enough that they dissociate more than at 2 °C. In other words, the rate of dissociation due to shear at 0 °C could be greater than that at 2 °C (for the shear rate used), such that there is a decrease in relative viscosity. However, this was not measured as part of this study, and further investigation is required to determine the effects in this temperature region.

### 3.2.2 Liquid to Solid Phase Transition

The effects of both temperature and pressure were measured in the same temperature range for concentrations of 0.1, 1, and 10 ppm O-MWCNT at pressures where hydrate formation did occur: 10 to 30 MPag. The viscosity-time behaviour of the successful runs is presented in **Figure 6** for the 1 ppm concentration. The figures for the 0.1 and 10 ppm systems are in the Appendix. The transition from the liquid-gas phase to the clathrate hydrate phase was characterized by increased viscosity over time until a maximum of about 1200 mPa·s was reached. At this point, nearly no liquid water remained. The transition also occurred in three stages: initial growth, the slurry phase, and then final growth. These stages were most distinct under the lowest driving force conditions: the blue 10 MPag or orange 15 MPag runs in **Figure 6**. A significant, unconstrained increase in viscosity was measured in the initial growth stage. The onset of the slurry phase was observed when the hydrate growth rate became limited, and viscosity became more stable. Note that slurry formation began at the onset of hydrate formation; it was made of suspended O-MWCNTs and hydrate clusters, and its length depended on the driving force for formation. Eventually, the final growth stage was reached, and there was a dramatic rise in viscosity until the maximum value was achieved.[29] These stages were observed for all test runs in which hydrates formed, though some driving forces were large enough that the slurry phase was undetectably short. From **Figure 6**, the temporal viscosity behaviour was correlated to temperature and pressure.

Specifically, hydrate formation occurred faster at lower temperatures and higher pressures (higher driving forces), reducing the length of the slurry phase and the time to the maximum viscosity. For instance, one can examine the limits of the horizontal axes for the 1 ppm system, which are seven times greater from 0 °C, where all runs were complete in under 10 minutes, to 8 °C, where all runs were complete in under 70 minutes. Therefore, the shortest phase transitions were observed at temperatures of 0 and 2 °C with pressures of 25 and 30 MPag, whereas the longest were observed at 8 and 10 °C with pressures of 15 MPag. Note that these same stages and behaviour were equally detected for the pure water-methane hydrate baseline (as well as in O-GNF-hydrate systems) and that the presence of O-MWCNTs may have only affected the stage time lengths. These kinetic effects will be discussed in section 3.2.3.

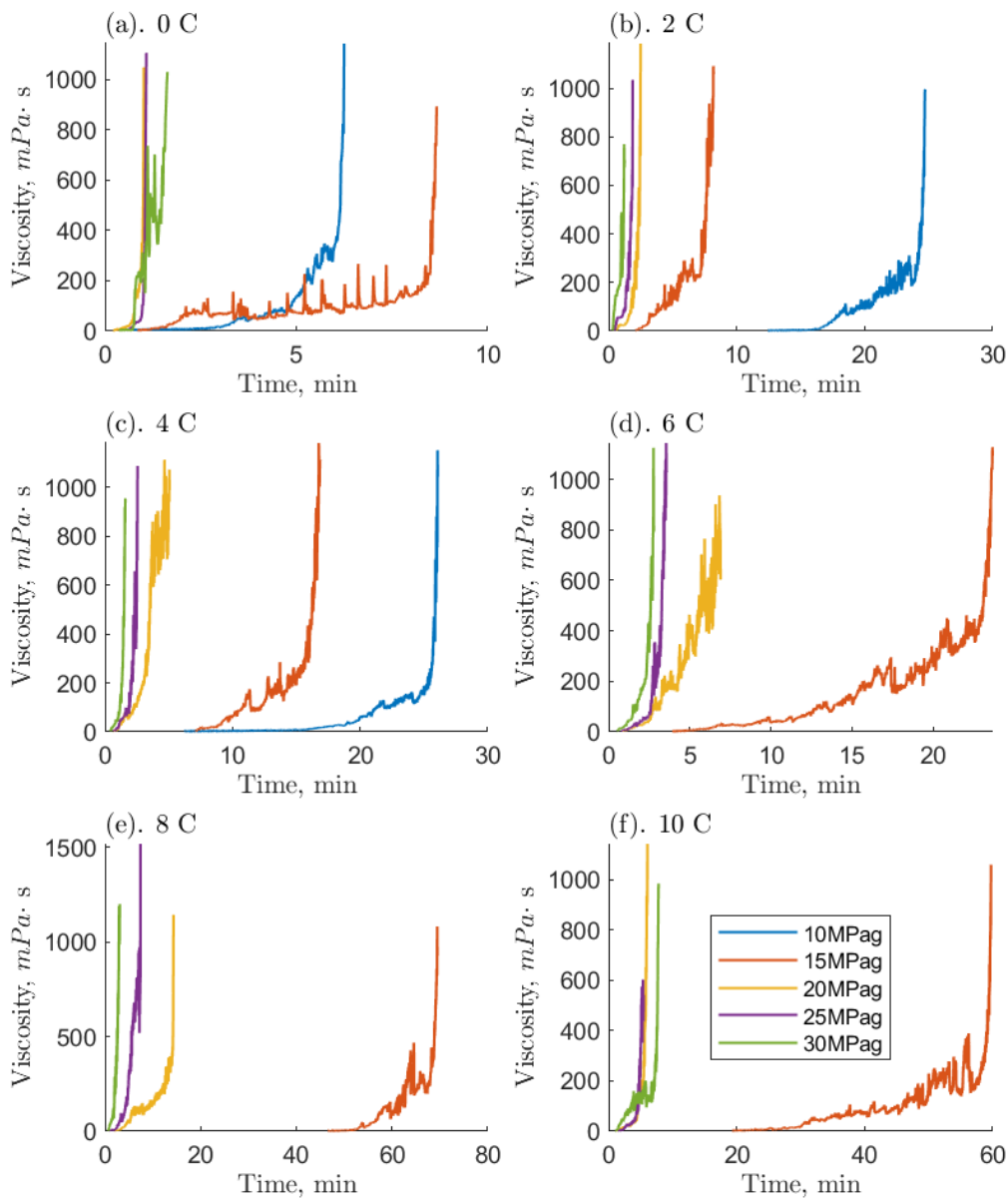

**Figure 6.** Measured temporal viscosities of the 1 ppm O-MWCNT-methane-water systems where hydrate formation occurred. Each subpanel (a-f) separates runs by temperature and contains isobaric viscosity time series starting at the onset of hydrate formation.

Compared to the baseline, systems with O-MWCNTs generally exhibited similar or faster times to reach the maximum viscosity. They also exhibited fewer prominent slurry phases, and those that were prominent tended to be shorter than their baseline counterparts. Note, however, that the 10 ppm system was unable to form hydrates at pressures of 10 MPag, so the comparison is mainly limited to the 0.1 and 1 ppm O-MWCNT systems. System limitations are discussed later in this section. O-MWCNTs have previously been described to promote hydrate formation in several ways. They can create localized fluid displacements by acting as microscopic stirrers, and increase the available gas-liquid interfacial area.[17] In addition, the shuttle effect could provide greater methane availability to the system. These effects can combine to enhance the mass transfer coefficient of the system, noting that mass transfer enhancement can be an order of magnitude more effective than heat transfer improvements for hydrate formation.[52, 53] Moreover, the motion of O-MWCNTs could increase the overall hydrate growth area by breaking up hydrate clusters. This would increase the hydrate growth rate and limit the length of the slurry phase. Lastly, mass diffusivity in the system could be enhanced by the hydrodynamic effect, where the nanoparticles collide and interact with the gas-liquid interface, which would induce turbulence and thin the effective diffusion layer.[54] Note that, in O-GNF systems, the times to the final viscosities were much faster than in the O-MWCNT systems. O-GNF systems also exhibited fewer, less prolonged slurry phases. This may be because of their higher specific surface areas, which would result in a higher gas-liquid interfacial area.

Examining the hydrate-forming runs qualitatively, the 0.1 and 1 ppm systems exhibited the least number of conditions with a significant and prolonged slurry phase despite the 10 ppm system forming hydrates in five fewer conditions. Furthermore, because of the smaller 10 ppm data set, the lengths of the slurry phases are only comparable between the 0.1 and 1 ppm systems and, examining them, the 0.1 ppm systems had somewhat faster slurry phase times. Previous studies by Pasieka et al. (2013) have demonstrated that methane hydrate growth rates were most enhanced at 0.1 ppm O-MWCNTs, followed by 10 and 1 ppm, the latter showing no enhancement due to a stronger increase in nanoparticle collisions with respect to the surface area.[17] This result is different from the current study, where 0.1 and 1 ppm are similar and faster than 10 ppm. However, in Pasieka et al. (2013), only the first 15 minutes of hydrate growth at a low driving force were used to determine the growth rate, limiting viscosity increases. When there are significant increases in viscosity, it is possible that the MWCNTs are not made part of the crystal matrix as more hydrate forms to solidification but are instead pushed further into solution, increasing the effective concentration in the liquid.[55] In the 1 ppm system, this effect could increase the amount of surface area per volume and overcome mean free path limitations, resulting in similar speeds compared to the 0.1 ppm system. Although, at 10 ppm, this same effect could also result in too high an effective concentration, leading to nanoparticle entanglement and reductions in surface area. The persistence length of these nanotubes is estimated to be between 270 and 420 nm, which means they can bend 20 to 40 times across their length.[16, 56] In turn, significant O-MWCNT entanglement would lead to decreases in hydrate growth enhancement. Observing slower rates at 10 ppm is similar to McElligott et al. (submitted), where the 10 ppm O-GNF system was also slowest of the three concentrations.[40] However, O-GNFs do not have the correct geometry to entangle, so the slowing was instead ascribed to mean free path limitations. Additionally, the 1 ppm O-GNF system was determined to be faster than the 0.1 ppm O-GNF system, rather than similar, due to surface area increases. This is likely because, due to their dimensions, O-GNFs have longer mean free paths compared to the larger, string-like O-MWCNTs: there may be some entanglement in the 1 ppm O-MWCNT system. Therefore, instead of the 1 ppm O-MWCNT system becoming faster than the 0.1 ppm system due to higher surface area, the two systems behave similarly. As concentration increases to 10 ppm, there is even more entanglement, and the system slows.

Notably, many of the driving force conditions under which hydrates could form did not successfully do so in the required 90-minute period. At 0.1 and 1 ppm O-MWCNT, hydrates formed successfully at driving forces at and above 6.2 MPa (4 °C/10 MPag): 11 positive driving force conditions did not form hydrates which would have in other systems.[11, 13] At 10 ppm O-MWCNT, hydrates successfully formed for driving forces at and above 10.3 MPa (6 °C/15 MPag), requiring over 4 MPa more than the other concentrations. Note that at 10 ppm, there was one successful formation event at 7.7 MPa (10 °C/15 MPag) though this was likely related to the stochasticity of nucleation and would not be reproducible. Therefore, 14 positive driving force conditions did not form hydrates. This is compared to the baseline, where hydrates formed for driving forces as low as 5.3 MPa, and O-GNF systems, which could form at 4.1 MPa. Certain limitations come with the specialized high-pressure rheological devices required to measure the properties of systems with gas hydrate formation. Namely, the shear environment, lack of nucleation surfaces, and diffusion limitations. The high-shear environment induced by the rheometer could cause mechanical dissociation of gas hydrate nuclei before they reach a critical radius. O-MWCNTs are long, string-like particles with low mean free paths; their presence increases the number of kinetic collisions in the solution and could lead to further dissociation of hydrate nuclei. Moreover, the smallest cross-section of an O-MWCNT is 774 nm$^2$, compared to 154 nm$^2$ for O-GNF (the largest are, respectively, 31,400 and 10,000 nm$^2$). Additional collisions between O-MWCNTs could generate a greater impact on the solution. Therefore, significantly higher driving forces for formation would be required for O-MWCNTs, explaining why hydrates form under fewer conditions in the nanotube systems. Furthermore, there are few impurities in the RO water, and the stainless-steel surfaces of the well and measurement system are quite smooth. Therefore, there are fewer sites for hydrate nucleation, noting that O-MWCNTs have been shown not to act as nucleation sites.[55] Finally, the small sample volume and double annulus measurement geometry result in a low gas-liquid surface area for diffusion, while heat evolution from hydrate formation could make the system self-limiting.[29] Carbon nanoparticles are expected to enhance hydrate systems, but instead, they do not affect the conditions under which hydrates form. Therefore, the 400 $s^{-1}$ shear rate, the only shear rate currently employed to measure viscosity in hydrate systems[29], may be too high, having the greatest limiting effect on the system. It is recommended for future work that different shear rates be tested to determine one that is optimal for forming hydrates: likely one that is lower, for example 300 $s^{-1}$.

Some of these limitations would also influence the slurry phase, though the length of this phase was shorter in O-MWCNT systems compared to the baseline. As mentioned, O-MWCNTs have previously been shown to enhance hydrate formation rates and methane dissolution rates in water by up to 16%.[17,18] Additionally, the slurry phase was generally more stable than the baseline: there were no significant drops in viscosity. These effects suggest that O-MWCNTs do not repress the inherent system limitations before hydrate formation (they may impose further limitations) but have some effect during growth. Improved mixing of the increased amount of methane from the presence of O-MWCNTs might have reduced mass-transfer limitations and allowed the system to maintain higher growth rates and viscosities during hydrate formation (if this formation occurred).

### 3.2.3 Methane Hydrate Growth Kinetics for Applications

To examine the growth rate behaviour as it pertains to the viscosity in hydrate technologies and give further insight into the phase transition, the times required to reach 200 ($T_{200}$) and 500 ($T_{500}$) mPa·s are presented in **Figure 7** for the 1 ppm system. The figures for the 0.1 and 10 ppm systems are in the Appendix. Any technological applications of hydrates will likely require limits to how viscosity can increase. This is to reduce the significant pumping requirements that often accompany higher viscosity values. Therefore, the time to reach the maximum, near-solid viscosity value is likely less pertinent to process or equipment design considerations. Moreover, the time required for the system to reach viscosities higher than 500 mPa·s was negligibly higher, and thus was omitted. From the figure, the fastest runs also had the highest driving forces, while those with the lowest driving forces were considerably slower. As the driving force increased, there was a non-linear decrease in the $T_{200}$ and $T_{500}$ values. The greatest decreases occurred between the 15 and 20 MPag runs, where times were cut approximately by at least two-thirds. This is compared to the values from 20 to 30 MPag, which exhibited little change despite the driving force increase. This behaviour is similar to the pure water baseline and was present at all concentrations.[29] In the baseline, all times at 20 MPag are relatively close, about 10 minutes separating the fastest and slowest times, which is the case for the 1 and 10 ppm O-MWCNT systems. However, in the 0.1 ppm system, the spread is about 35 minutes at 20 MPag (though the 25 and 30 MPag times remain similar). It is optimal from an application standpoint to have similar times across temperatures and pressures so that the same hydrate growth rates can be achieved for less severe conditions. For instance, in the 1 ppm system, the $T_{500}$ values at 8 °C/25 MPag and 6 °C/20 MPag are 5.53 and 5.57 minutes, respectively. This corresponds to much less severe conditions to reach nearly the same timescale, which could indicate that the system's limits are possibly being approached and that 0.1 ppm is not an optimal concentration compared to 1 ppm. However, similar times do not necessarily mean the system is faster than the baseline, as the same aggregation kinetics exist across multiple conditions. This could result in less significant variations in $T_{200}$ and $T_{500}$ values at either 1 or 10 ppm. Moreover, the O-GNF systems were much faster than the O-MWCNT ones: there was essentially no change from 10 to 30 MPag, relative to O-MWCNT values, and the O-GNF curves would appear as flat lines if plotted in **Figure 7**.

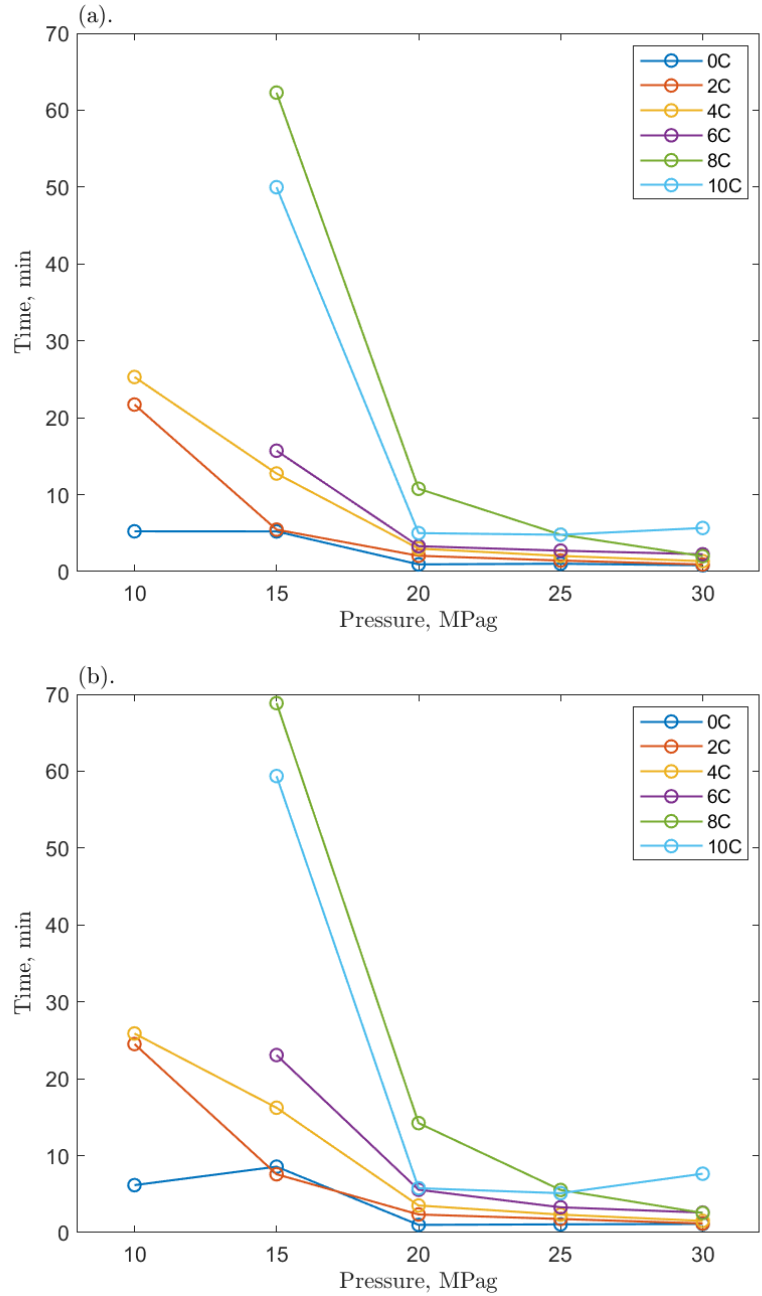

**Figure 7.** The time required for the 1 ppm O-MWCNT-methane-water system to reach (a) 200 mPa·s and (b) 500 mPa·s from the onset of hydrate formation at various pressures.

Compared to the baseline $T_{200}$ and $T_{500}$ values, those for the O-MWCNT systems were increased, demonstrating a slower initial growth rate despite reaching the maximum viscosity value faster. No clear trends for the change from the baseline could be determined between individual values across any of the temperatures, pressures, or concentrations. Only general, average values of the change across all conditions for each concentration showed apparent effects. However, because these are averages between many temperature/pressure conditions, the significance of these values may be minimal. The lack of notable trend is possibly due to the short time frames in this study (some runs were complete in just a few seconds) and the high stochasticity of the nucleation process. The $T_{200}$ times were increased by 81.29, 28.02, and 28.30 % on average for the 0.1, 1, and 10 ppm systems, respectively. The $T_{500}$ times were increased by 94.6, 21.08, and 34.02 % on average for the 0.1, 1, and 10 ppm systems, respectively. These are slower than the baseline and much slower than O-GNF systems, which lowered the T200 and T500 values maximally by 49.75 and 31.92 %, respectively, at 1 ppm O-GNF. This may be because the slower slurry phase usually occurs at viscosities above 500 mPa·s in pure water and O-GNF systems, whereas, because of entanglement effects upon hydrate formation in O-MWCNT systems, the slurry phase occurs at lower viscosities, between 200 and 500 mPa·s. While the final viscosity times are faster than the baseline, the system is initially slower to reach critical viscosity values. This earlier slowdown is not optimal for technological applications as it may make viscosity more challenging to control due to slower response rates.

However, the $T_{200}$ and $T_{500}$ values at higher driving force conditions show more similarities between the two nanoparticle systems. One could take the average of the values only from 20 to 30 MPag, eliminating the influence of the significantly slower runs at the lowest driving forces and equalizing the number of values in each data set being compared. Notably, when this average is taken, the 1 ppm O-MWCNT system is now faster than the baseline ($T_{200}$ and $T_{500}$ values are decreased by 22.86 and 25.39 %, respectively), and the 10 ppm O-MWCNT system has no statistically significant change in growth rate compared to the baseline. While these are still not faster than O-GNF systems, these averages demonstrate that there are conditions where the O-MWCNT systems improve on baseline growth rate values and that the start of the slurry phase at lower viscosities occurs mainly at lower driving forces, likely as there is more time for nanotube entanglement under those conditions.[55] Specifically, the 1 ppm O-MWCNT system now enhances the system, whereas there is no significant enhancement from the 10 ppm system, and at 0.1 ppm, the times are still much slower than the baseline. This may occur because the 1 ppm condition has the highest surface area for the least amount of entanglement. The 0.1 ppm system may not enhance the system sufficiently to overcome the early slurry phase, while more entanglement or aggregation in the 10 ppm system limits surface area improvements. Therefore, in applications with significant viscosity changes, O-MWCNT nanofluids may only be useful at high driving forces and, compared to O-GNFs, could require concentrations that are specific to a chosen viscosity. These conclusions outline the importance of this work for exploring and elucidating the design considerations relevant to hydrate-based technologies.

## 4. CONCLUSIONS

The viscosity of O-MWCNT nanofluids was measured for concentrations from 0.1 to 10 ppm under conditions of 0 to 30 MPag pressures and 0 to 10 °C temperatures. This was the first time that the viscosity of plasma-functionalized carbon nanotubes had been measured in liquid, high-pressure, or hydrate-forming systems. Measuring the viscosity of these nanoparticles at the specified concentrations was also novel. The presence of O-MWCNTs did not affect the temperature dependence of viscosity in water. However, when added to water, the effective viscosity of solution was reduced from 0.1 to 1 ppm, though an increase would be expected. This non-Einsteinian behaviour may have been due to reductions in hydrogen bond strength at the hydrophobic portion of the O-MWCNT surface and enhanced density fluctuations at that surface. Together, these increase the number of larger, empty sites (greater free volume) and lower the impedance for water to diffuse to those sites. These surface effects may overcome internal friction that would otherwise raise viscosity, as the concentrations were ultra-low though the nanoparticle surface area remained high. This hypothesis was not tested in this study and is suggested for future work using computational models.

The addition of O-MWCNTs resulted in the creation of a weak, negative pressure dependence of viscosity in water. This may have resulted from a greater alignment of the nanotubes with the flow direction with increased pressure (i.e., the more aligned the particles were, the less drag they would add to the system, and viscosity would be slightly reduced). When pressurized, however, the system's relative viscosity was largely non-Einsteinian except at 2 °C. This may have been related to the density anomaly of water which was moved down from 4 °C due to weaker hydrogen bonds and the balance between the formation and dissociation of hydrate nuclei and ice-like structures in the shear environment. The liquid to solid (hydrate) phase transition was divided into initial growth, a slurry phase, and final growth to a maximum viscosity. The times to reach that viscosity were faster in O-MWCNT systems than in the baseline. They also exhibited shorter slurry phase times due to enhanced mass transfer. The 0.1 and 1 ppm systems were equally fast, while the 10 ppm system was slower. This was possibly due to greater nanotube entanglement at the higher concentration, which was exacerbated by the growth of a hydrate phase that pushed the O-MWCNTs further into the liquid and resulted in an effectively higher concentration. The presence of O-MWCNTs did not overcome the limits to hydrate formation present in the baseline study and formed hydrates under even fewer conditions. This may have been because their short mean free paths resulted in greater collisions in the system and inhibited the formation of critical clusters of hydrate nuclei. The times to viscosity values most relevant to technological applications were minimally 28.02 % (200 mPa·s) and 21.08 % (500 mPa·s) slower than the baseline, both in the 1 ppm system, even though the system was faster to the final viscosity value. This was because the slurry phase occurred at much lower viscosities and was particularly long at lower driving forces. Using only higher driving force runs (pressures of 20 MPag and above), the system at 1 ppm showed faster $T_{200}$ and $T_{500}$ values (by 22.86 and 25.39 %, respectively), indicating that O-MWCNTs may only be useful in hydrate systems at higher driving forces.

Compared to O-GNFs, O-MWCNTs had relative viscosities closer to 1.00 or above 1.00 at higher (5 to 10 ppm) concentrations. This was likely because O-GNFs have greater specific surface areas, so they may exhibit a more significant accumulation of viscosity-reducing surface effects. They also have higher mean free paths, so there were fewer collisions in the system. Furthermore, O-GNF systems were faster to the final viscosity while exhibiting less significant slurry phase behaviour and formed hydrates at more driving force conditions. Finally, O-GNFs were faster to critical viscosity values at all concentrations, again due to enhanced mass transfer effects. These results outline the importance of the specific surface area, geometry, and dispersion when evaluating the efficacy of additives in hydrate-forming technological systems.

AUTHOR INFORMATION

**Corresponding Author**

*phillip.servio@mcgill.ca

**Author Contributions**

The manuscript was written through the contributions of all authors. All authors have given approval to the final version of the manuscript.

**Conflicts of Interest**

There are no conflicts to declare.

ACKNOWLEDGEMENTS

The authors would like to acknowledge the financial support from the Natural Sciences and Engineering Research Council of Canada (NSERC) and the Faculty of Engineering of McGill University (MEDA, Vadasz Scholars Program).

APPENDIX

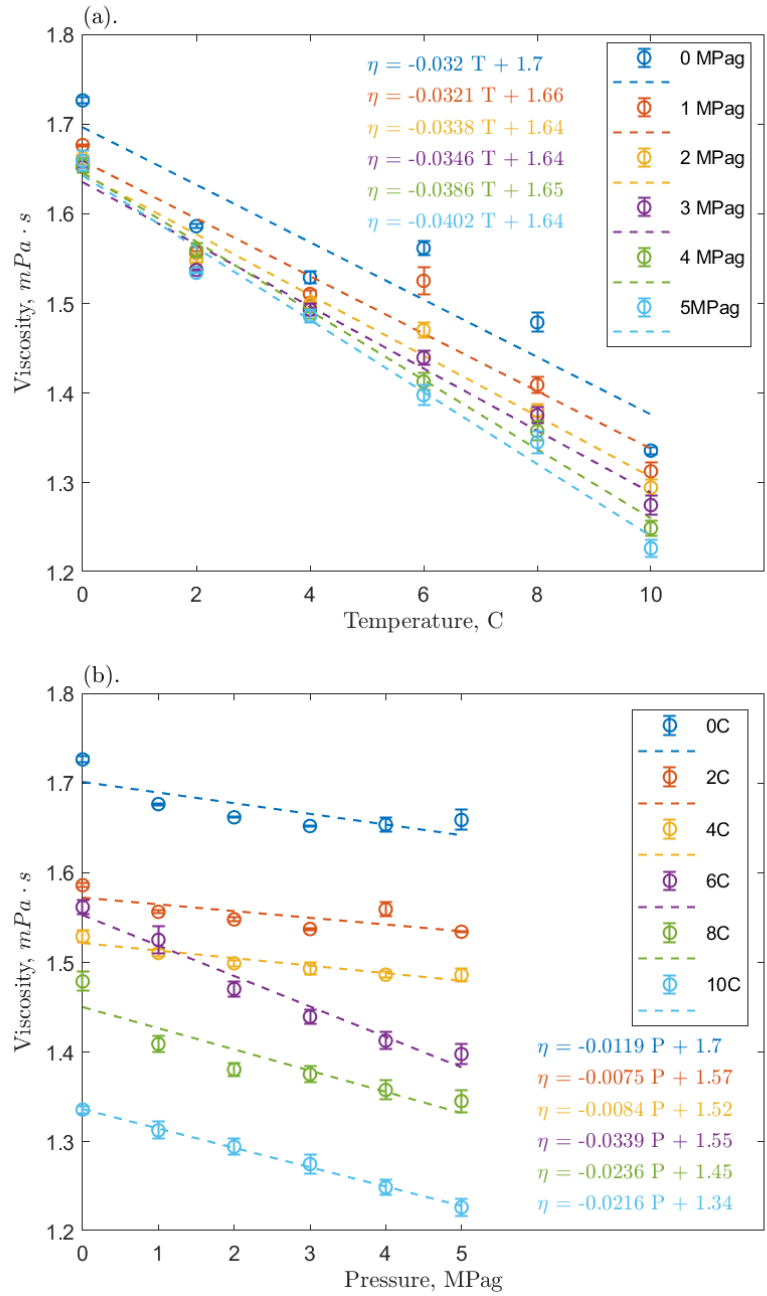

**Figure A1.** Effect on the measured viscosity of the 0.1 ppm O-MWCNT-methane-water system of (a) temperature and (b) pressure; error bars represent the 95% confidence intervals on the measured mean viscosity.

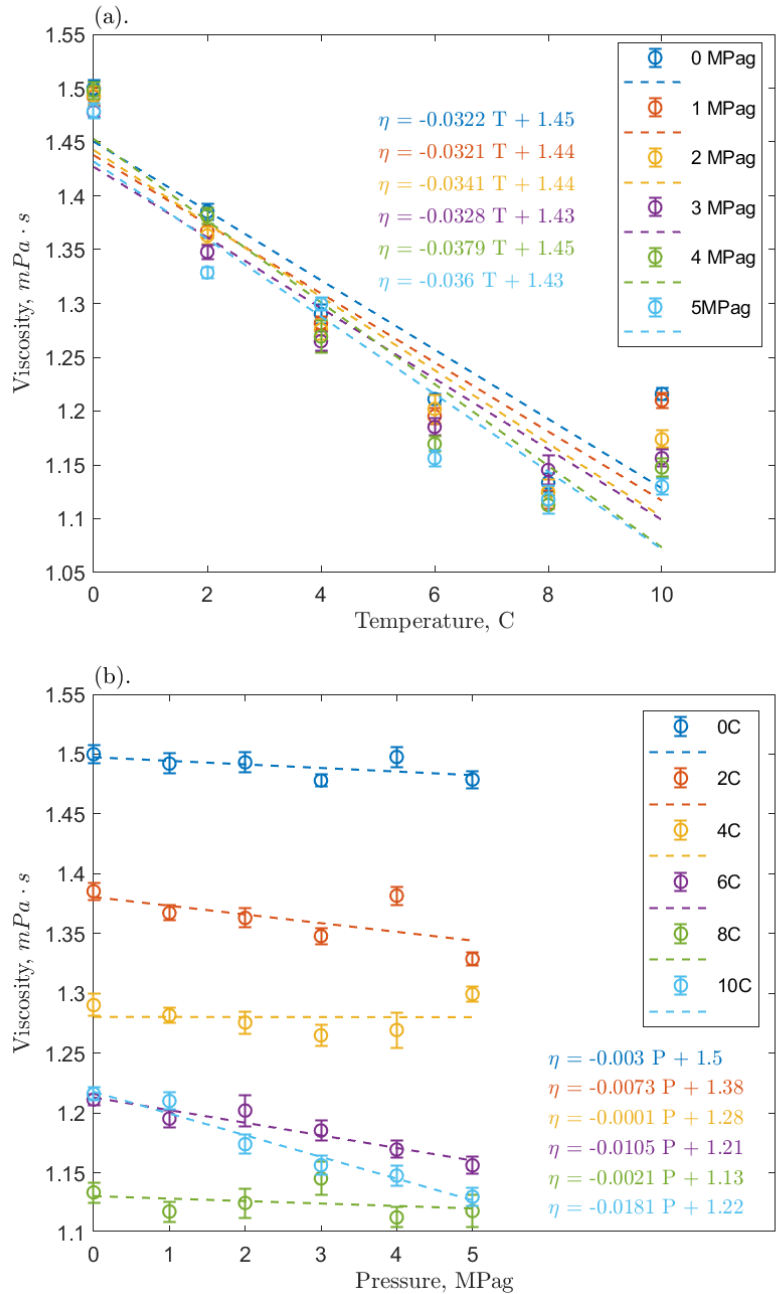

**Figure A2**. Effect on the measured viscosity of the 10 ppm O-MWCNT-methane-water system of (a) temperature and (b) pressure; error bars represent the 95% confidence intervals on the measured mean viscosity.

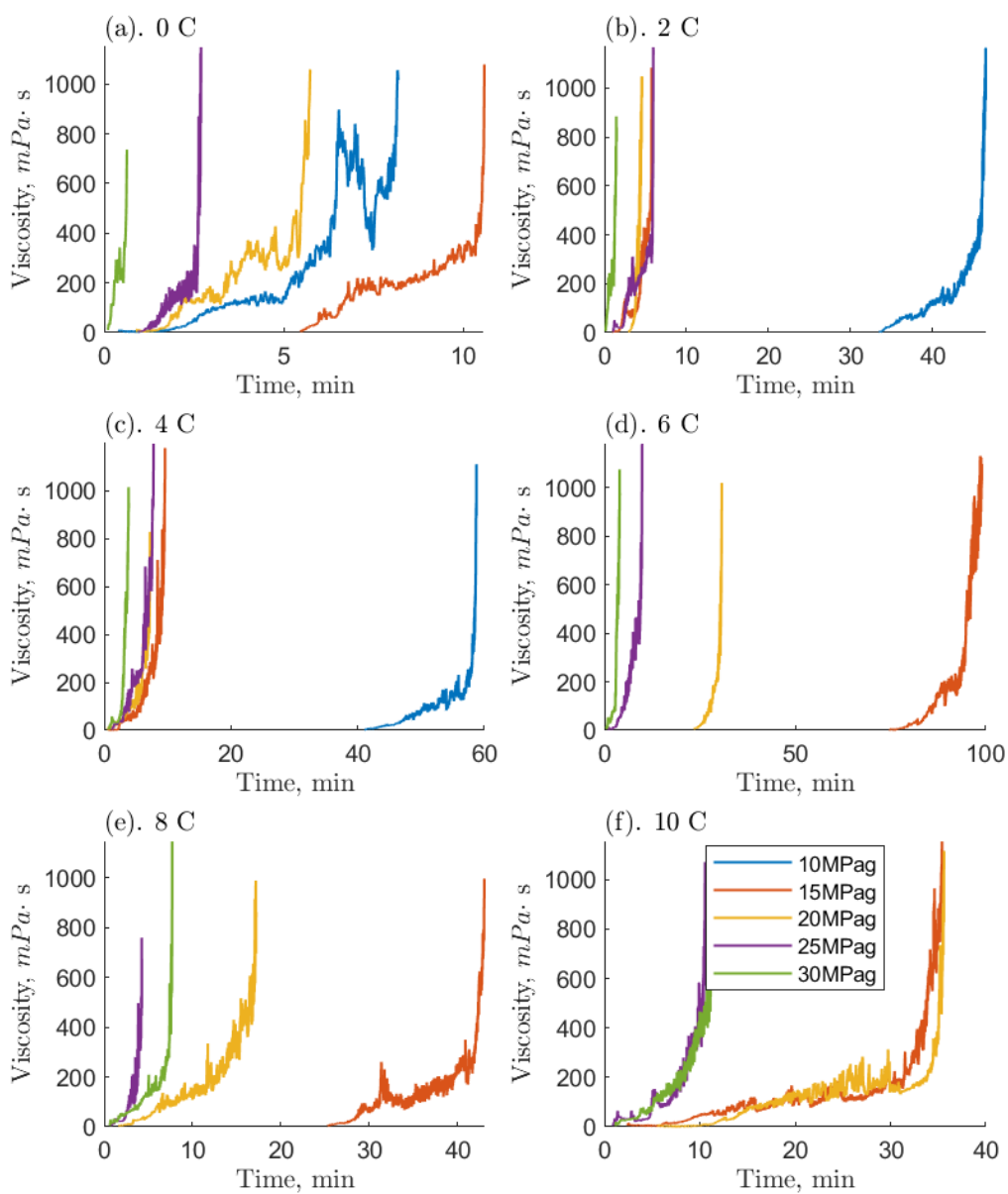

**Figure A3.** Measured temporal viscosities of the 0.1 ppm O-MWCNT-methane-water systems where hydrate formation occurred. Each subpanel (a-f) separates runs by temperature and contains isobaric viscosity time series starting at the onset of hydrate formation.

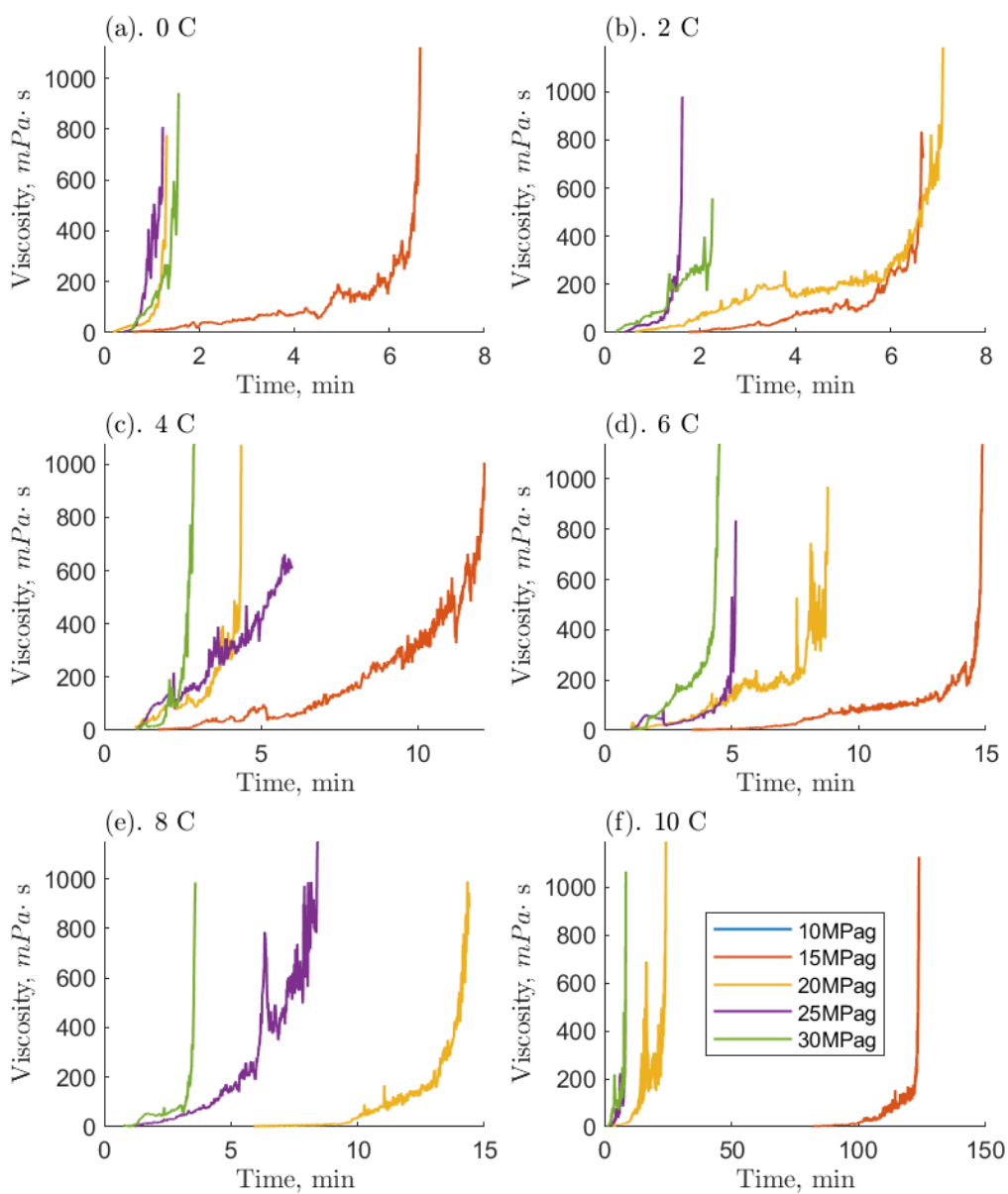

**Figure A4.** Measured temporal viscosities of the 10 ppm O-MWCNT-methane-water systems where hydrate formation occurred. Each subpanel (a-f) separates runs by temperature and contains isobaric viscosity time series starting at the onset of hydrate formation.

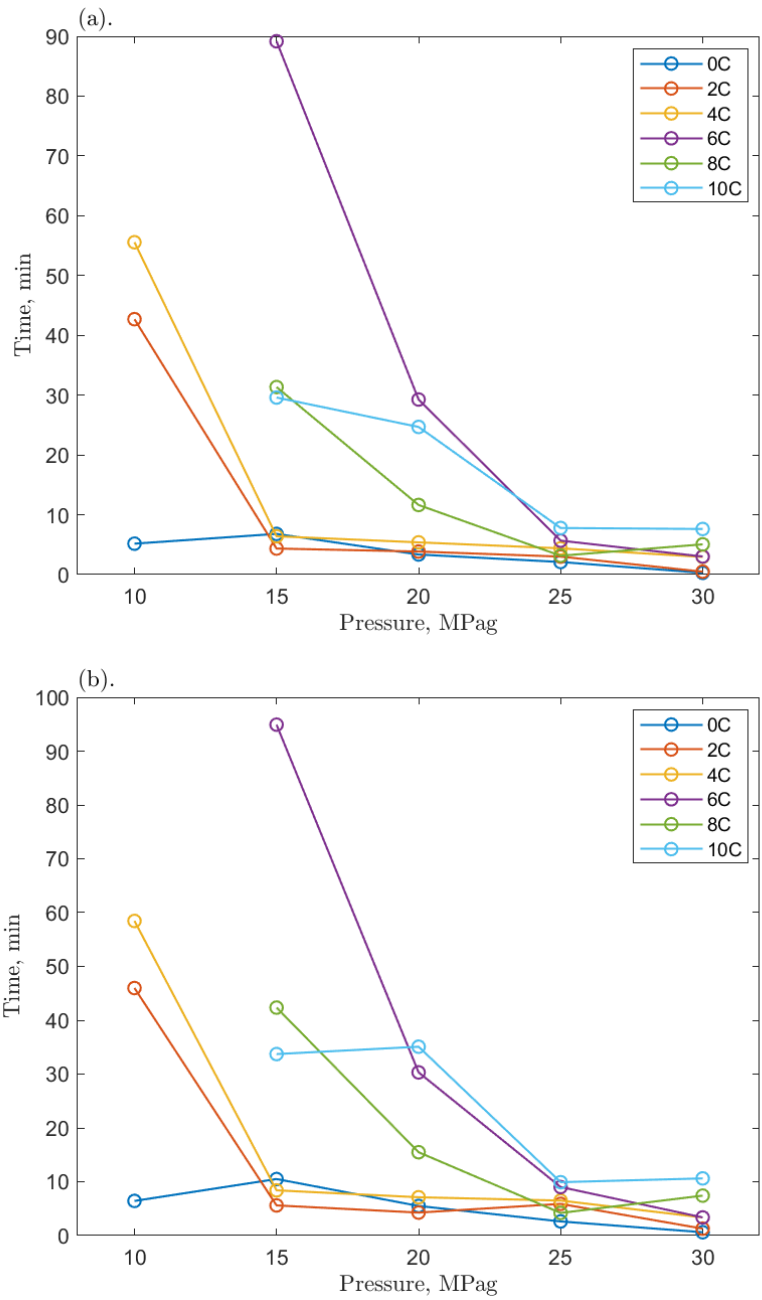

**Figure A5.** The time required for the 0.1 ppm O-MWCNT-methane-water system to reach (a) 200 mPa·s and (b) 500 mPa·s from the onset of hydrate formation at various pressures.

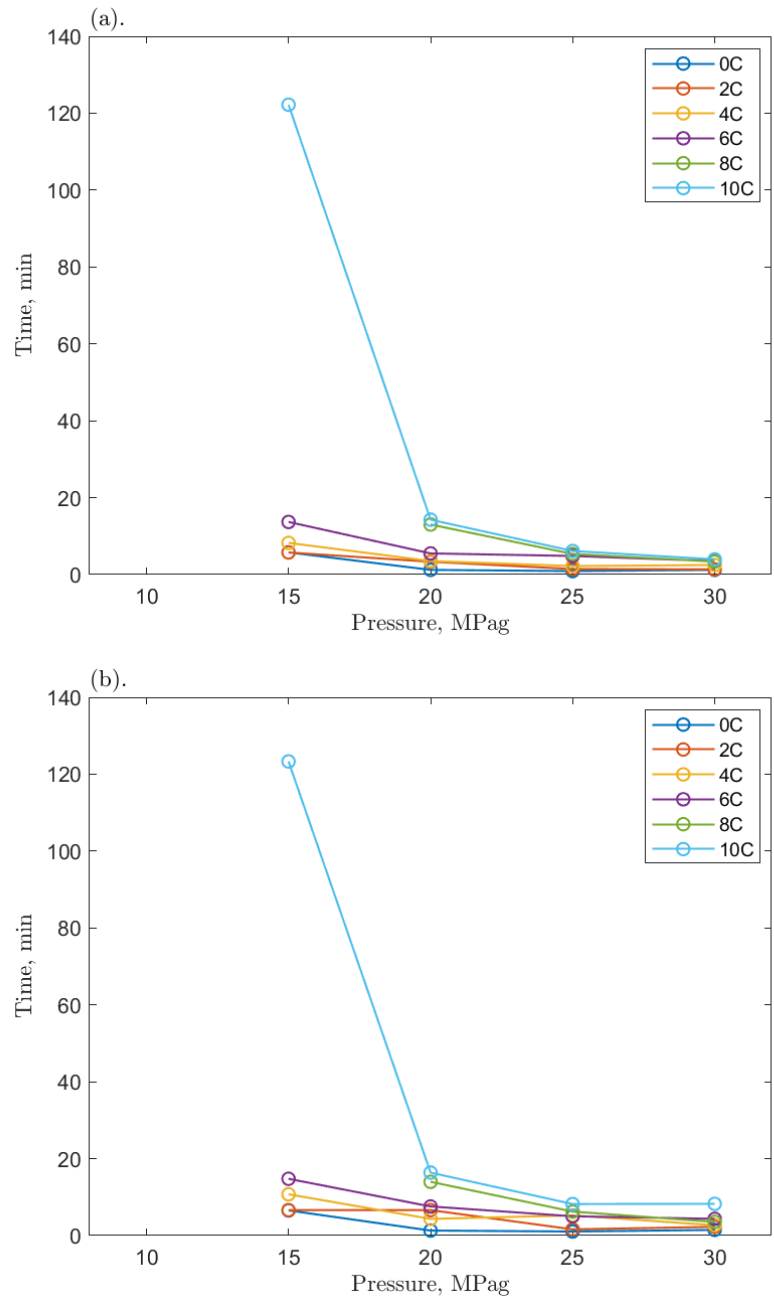

**Figure A6.** The time required for the 10 ppm O-MWCNT-methane-water system to reach (a) 200 mPa·s and (b) 500 mPa·s from the onset of hydrate formation at various pressures.